\documentstyle[amsfonts,12pt]{article}
\begin{document}
\title{Intertwining relations of non-stationary Schr\"odinger operators}
\author{
F.\ Cannata$^*$\thanks{E-mail: cannata@bo.infn.it},
M.\ Ioffe$^\dagger$\thanks{E-mail: ioffe@snoopy.phys.spbu.ru},
G.\ Junker$^\ddagger$\thanks{E-mail: junker@theorie1.physik.uni-erlangen.de}
and 
D.\ Nishnianidze$^\dagger$\thanks{E-mail: david@heps.phys.spbu.ru}\\[5mm]
$^*$ Dipartimento di Fisica and INFN, Via Irnerio 46, 40126 Bologna,
Italy\\[2mm] 
$^\dagger$ Department of Theoretical Physics, University of St.\ Petersburg,\\
198904 St.\ Petersburg, Russia\\[2mm]
$^\ddagger$ Institut f\"ur Theoretische Physik, Universit\"at
Erlangen-N\"urnberg,\\
 Staudtstr.\ 7, D-91058 Erlangen, Germany}
\date{}
\maketitle
\begin{abstract}
General first- and higher-order intertwining relations between non-stationary
one-dimensional Schr\"odinger operators are introduced. For the first-order
case it is shown that the intertwining relations imply some hidden
symmetry which in turn results in a $R$-separation of variables. The
Fokker-Planck and diffusion equation are briefly considered. Second-order
intertwining operators are also discussed within a general approach. However,
due to its complicated structure only particular solutions are given in some
detail. 
\end{abstract}

\section{Introduction}
It is a basic fact that many physical phenomena are mathematically
described by solutions of linear ordinary and partial differential
equations. For example, the dynamics of a classical system may be
characterized by Newton's equation or equivalently by the Euler-Lagrange
equation. Another example is the dynamics of a non-relativistic quantum system
which is governed by a linear differential equation, the well-known
Schr\"odinger equation. Hence, exact solutions of such kind of linear
differential equations are of basic interest and, 
therefore, much effort has been made during the last centuries to find
solutions of problems being of the form
\begin{equation}
  {\cal L}\psi=0\;,
\label{Lpsi}
\end{equation}
where ${\cal L}$ denotes some linear differential operator. Here $\psi$ is the
wanted solution of (\ref{Lpsi}) in some function space associated with given
initial and/or boundary conditions. Many methods have been developed to find
such solutions. This paper is concerned with one of them based on an
assumed {\em intertwining relation}
\begin{equation}
  {\cal L}_1{\cal I}={\cal I}{\cal L}_2
\label{inter}
\end{equation}
between two physically relevant differential operators ${\cal L}_1$, 
${\cal L}_2$ and the so-called {\em intertwining operator} ${\cal I}$, which
is also assumed to act linearly. This intertwining relation allows to
construct a solution for, 
say, ${\cal L}_1$ if a solution of ${\cal L}_2$ is known. To be more precise,
let us assume that $\psi_2$ is a solution of ${\cal L}_2\psi_2=0$, then due to
the intertwining relation $\psi_1={\cal I}\psi_2$ is a solution of ${\cal
  L}_1\psi_1=0$. 
Note that in addition we have to assume that $\psi_2$ does not belong to the
kernel of ${\cal I}$, i.e.\ ${\cal I}\psi_2\not\equiv 0$. 
Multiplying (\ref{inter}) from left and right with the inverse operator ${\cal
  I}^{-1}$ we have ${\cal I}^{-1}{\cal L}_1={\cal L}_2{\cal I}^{-1}$ and,
therefore, one can also start with a given solution $\psi_1$ and obtain a new
solution $\psi_2={\cal I}^{-1}\psi_1\not\equiv 0$. However, in many cases
the intertwining operator is a differential operator. Therefore, its inverse
is a rather complicated integral operator and of less use. 
This situation, however, changes if
in addition we assume that both operators ${\cal L}_{1,2}$ are self-adjoint on
some common function space, ${\cal L}_{1,2}^\dagger={\cal L}_{1,2}$. As a
consequence one has the adjoint intertwining relation 
\begin{equation}
  {\cal I}^{\dagger}{\cal L}_1={\cal L}_2{\cal I}^{\dagger}\;,
\label{inter+}
\end{equation}
which in addition provides us with the symmetry operators ${\cal I}{\cal
  I}^{\dagger}$ and ${\cal I}^{\dagger}{\cal I}$ obeying \cite{ain1-3}
\begin{equation}
  [{\cal L}_1,{\cal I}{\cal I}^{\dagger}]=0\;,\qquad
  [{\cal L}_2,{\cal I}^{\dagger}{\cal I}]=0\;.
\label{symmetry}
\end{equation}

The above connection between solutions of two differential equations have, to
our knowledge, been applied first by 
Darboux \cite{Darboux82} to Sturm-Liouville problems. That is, ${\cal
  L}_{1,2}$ belong to a class of second-order differential operators and may,
for example, represent the Hamiltonian of a quantum mechanical degree of
freedom in one dimension. This fact may be the reason why most applications of
such intertwining relations (sometimes also called Darboux transformations)
have been made for eigenvalue problems of stationary Schr\"odinger
Hamiltonians in one dimension. Here, ${\cal L}_{1,2}$ represent a pair of
Schr\"odinger Hamiltonians intertwined by a 
first-order time-independent differential operator. This pair of Hamiltonians
together with the intertwining operator (also called supercharge) form the
basis of Witten's model of supersymmetric (SUSY) quantum mechanics
\cite{Junker96}. It is also closely related to the factorization method of 
Schr\"odinger \cite{Schroedinger} and Infeld and Hull \cite{Infeld}, and
still finds application in the construction of new potentials giving rise to
exactly solvable Schr\"odinger-eigenvalue problems. See, for example,
\cite{Junker98,Cannata98}. This approach has also been generalized
\cite{abi84,abei85} to higher-dimensional systems where matrix Hamiltonians
necessarily participate in the intertwining relations. Very recently,
intertwining relations with supercharges of second (and higher) order in
derivatives were investigated for one-dimensional \cite{ais,acdi} and
two-dimensional systems \cite{ain1-3} as well as for matrix Hamiltonians in
one dimension \cite{acin}. In the one-dimensional case it was shown that
irreducible second-order transformations, these are those which cannot be
represented by two consecutive standard (first-order) Darboux transformations,
do exist. In two dimensions second-order irreducible transformations allow to
avoid intermediate matrix Hamiltonians. Both, for one-dimensional matrix and
two-dimensional scalar quantum systems non-trivial symmetry operators
(\ref{symmetry}) were constructed \cite{ain1-3,acin} and led to integrability
of the corresponding dynamical systems.

In this paper we will consider intertwining relations of the {\em
  non-stationary Schr\"odinger operator}
\begin{equation}
 S[V]=i\partial_t+\partial^2_x - V(x,t)\;.
  \label{S}
\end{equation}
Here $\partial_t=\partial/\partial t$ and 
$\partial_x=\partial/\partial x$ denote the partial derivatives with respect
to time and position. Throughout this paper we will denote these derivatives,
if applied to some function $f$, by a dot and prime, respectively. Hence, we
use the notation $\dot{f}(x,t)=(\partial_t f)(x,t)$ and $f'(x,t)=(\partial_x
f)(x,t)$. It is also evident that we are going to use units such that Planck's
constant $2\pi\hbar$ and the mass $m$ are given by $\hbar=2m=1$.

Replacing the general operator ${\cal L}$ by the Schr\"odinger operator
(\ref{S}) the corresponding equation (\ref{Lpsi}) reads $S[V]\psi(x,t)=0 $ and
represents the non-stationary Schr\"odinger equation for a one-dimensional
quantum system under the influence of an external time-dependent potential 
$V:{\mathbb   R}\times{\mathbb R}\to{\mathbb   R},(x,t)\mapsto V(x,t)$. A
first investigation of this type of 
intertwining relations is due to Bagrov and Samsonov \cite{Bagrov96} who
considered a time-dependent first-order intertwining operator. See also the
review \cite{Bagrov97} where in addition a reducible second-order
intertwining of non-stationary Schr\"odinger operators is considered.

In the next section we will reconsider the first-order intertwining and show
that, due to the $R$-separation of variables, it provides no basic new results
besides those already known from the 
stationary case of time-independent potentials. This observation has also
recently been anticipated by Finkel et al.\ \cite{Finkel98}. However, we
additionally show that this fact can be related to some underlying symmetry
structure. We will 
also briefly discuss the time-dependent Fokker-Planck equation in this context
and find that this case does not lead to a trivial result in the above
sense. In Section 3 we will present particular solutions for stationary
potentials intertwined by a non-stationary, i.e.\ time-dependent, second-order
operator. We show that the reducible cases (as studied in \cite{Bagrov97}) are
only very special cases of this class. In Section 4 we finally discuss the
intertwining of non-stationary Schr\"odinger operators by non-stationary
second-order operators. As special cases we consider the intertwining of the
Schr\"odinger operators corresponding to a non-stationary Hamiltonian with
that for a stationary one as well as with that of a time-dependent harmonic
oscillator. 
In the concluding remarks in Section 5 we discuss some
aspects omitted in the main text such as zero modes, domain question and
generalizations to complex potentials.

\section{Intertwining by first-order operators}
In this section we will consider the most general first-order intertwining
relations for the Schr\"odinger operator as defined in (\ref{S}) and the
Fokker-Planck operator $F[U]$,
\begin{equation}
\label{F}
   F[U]=-\partial_t+\partial^2_x +\partial_x U'(x,t)
       =-\partial_t+\partial^2_x + U'(x,t)\partial_x + U''(x,t)
   \;.
\end{equation}
In the latter case, the relation (\ref{Lpsi}) becomes the well-known
Fokker-Planck equation \cite{Kampen92} 
characterizing the diffusion of a macroscopic (one-dimensional) degree of
freedom in an external time-dependent drift potential 
$U:{\mathbb R}\times{\mathbb R}\to{\mathbb R}, (x,t)\mapsto U(x,t)$ 
with a diffusion constant set equal to unity. For both cases we start with the
most general intertwining operator of first order given by 
\begin{equation}
  \label{q+}
  q^+(x,t)=\xi_0(x,t)\partial_t + \xi_1(x,t)\partial_x + \xi_2(x,t)
\end{equation}
with, in general, complex-valued functions $\xi_0,\xi_1$ and $\xi_2$. Note
that in contrast to \cite{Bagrov96} we also allow for a first-order operator
in $\partial_t$.

\subsection{The Schr\"odinger operator and separation of variables}
For the above defined Schr\"odinger operator (\ref{S}) the most general
first-order intertwining relation reads
\begin{equation}
  \label{inS}
  S[V_1]q^+ = q^+ S[V_2]\;,
\end{equation}
where the functions $\xi_i$ ($i=0,1,2$) and the potentials $V_{1,2}$
are clearly not independent of each other. In fact, we will now consider the
question: Which $\xi$'s lead to real-valued potentials and how are the latter
connected with the them?

Inserting the explicit forms of the Schr\"odinger operators (\ref{S}) and the
intertwining operator (\ref{q+}) into relation (\ref{inS}) and equating the
coefficients 
of identical (partial) differential operators we immediately find that $\xi_0$
and $\xi_1$ may only depend on time, i.e.\ $\xi_0'=0=\xi_1'$. The assumption
that $\xi_0$ does not vanish identically then leads to the conclusion that the
potential difference $V_1-V_2$ does depend only on time. This is a rather
trivial case\footnote{Note that from $V_2(x,t)-V_1(x,t)=f(t)$ and
  $S[V_1]\psi_1=0$ follows that $S[V_2]\psi_2\equiv S[V_1+f]\psi_2=0$ has the
  solution $\psi_2(x,t)=\exp\{i\int_0^td\tau\,f(\tau)\}\psi_1(x,t)$. That is,
  knowing a normalized solution of the time-dependent Schr\"odinger equation
  for $V_1$ one immediately has a solution for $V_2$, too, which is also
  normalized. Complex-valued $f$'s are not allowed as both, $V_1$ and $V_2$,
  are assumed to be real-valued.\label{footnote1}} 
and, therefore, we will set $\xi_0\equiv 0$ without loss of
generality.\footnote{A posteriori, this justifies the ansatz made in
  \cite{Bagrov96}.} 
Making now the following reparametrizations $\xi_1(t)=e^{i\beta(t)}\rho(t)$ and
$\xi_2(x,t)=e^{i\beta(t)}\rho(t)\chi'(x,t)$ with $\beta:{\mathbb R}\to{\mathbb
R}$, $\rho:{\mathbb R}\to{\mathbb R}^+$ and $\chi:{\mathbb R}\times{\mathbb
R}\to {\mathbb C}$ we find
\begin{equation}
  \label{V1andV2:1}
  \begin{array}{l}
   V_1(x,t)=\chi'\,^2(x,t)+\chi''(x,t) - i \dot{\chi}(x,t) +
   \alpha(t)-\dot{\beta}(t)+i\dot{\rho}(t)/\rho(t)\;,\\[2mm]
   V_2(x,t)=\chi'\,^2(x,t)-\chi''(x,t) - i \dot{\chi}(x,t) + \alpha(t)\;,
  \end{array}
\end{equation}
where $\alpha$ is some time-dependent complex-valued integration constant. 
Following the argumentation made in footnote \ref{footnote1} we may set
$\beta\equiv 0$ without loss of generality. Furthermore, we may also set
$\alpha\equiv 0$ because it can always be absorbed in $\chi$ via the
substitution $\chi\to \chi -i\int dt\,\alpha$. Hence, we are left with 
\begin{equation}
  \label{V1andV2:2}
  \begin{array}{l}
   V_1(x,t)=\chi'\,^2(x,t)+\chi''(x,t) - i \dot{\chi}(x,t) +
   i\dot{\rho}(t)/\rho(t)\;,\\[2mm]
   V_2(x,t)=\chi'\,^2(x,t)-\chi''(x,t) - i \dot{\chi}(x,t)\;.
  \end{array}
\end{equation}
Here the so-called superpotential $\chi$ is still not arbitrary as the
potentials are assumed to be real-valued. This can, for example, be achieved
by assuming a stationary real-valued superpotential. This, however, leads to
the standard stationary supersymmetric quantum mechanics discussed to a large
extend during the last two decades \cite{Junker96}. Therefore, we will
consider a complex-valued superpotential 
\begin{equation}
  \chi(x,t)= h(x,t) + i g(x,t)
\label{chi}
\end{equation}
with real-valued functions $h$ and $g\not\equiv 0$. The reality condition
  $\mbox{Im }V_1=\mbox{Im }V_2=0$ then leads to
\begin{equation}
  \label{h_and_g}
  2g''+\dot{\rho}/\rho=0\;,\qquad 2h'g'-g''-\dot{h}=0\;,
\end{equation}
which can easily be integrated to
\begin{equation}
  \begin{array}{rcl}
g(x,t)&=&\displaystyle
      -\frac{1}{4}\,\frac{\dot{\rho}(t)}{\rho(t)}\,x^2+
            \frac{1}{2}\,\rho(t)\dot{\mu}(t)x +\gamma(t)\;, \\
  h(x,t)&=&\displaystyle
      \frac{1}{2}\,\ln\rho(t)+K\Bigl(x/\rho(t)+\mu(t)\Bigr)\;,
  \end{array}\label{g_and_h}
\end{equation}
where $\mu$ and $\gamma$ are arbitrary real-valued function of time and $K$ is
an arbitrary real-valued function of the variable $y=x/\rho +\mu$. In terms of
these functions the final form of the two partner potentials is
\begin{equation}
  V_{1,2}(x,t)=\frac{1}{\rho^{2}(t)}\left[K'^2(y)\pm K''(y)\right]-
  \frac{\ddot{\rho}(t)}{4\rho (t)}\,x^{2}+\left(\dot{\rho}(t)\dot{\mu}(t)+
  \frac{\rho(t)\ddot{\mu}(t)}{2}\right)x-\frac{\rho^2(t)\dot{\mu}^2(t)}{4}
  +\dot{\gamma}(t) 
\label{V_1/2}
\end{equation}
and the intertwining operator reads
\begin{equation}
  q^+(x,t)=\rho(t)\partial_x+K'\Bigl(x/\rho(t)+\mu(t)\Bigr)-
  \frac{i}{2}\left(\dot{\rho}(t)x-\rho^2(t)\dot{\mu}(t)\right).
\label{q+_2}
\end{equation}
It is obvious that with appropriate choices for $K$, $\rho$, $\mu$  and
$\gamma$ at least one 
of the two potentials (\ref{V_1/2}) can be made stationary. We will not
investigate this aspect here but rather like to discuss a more general
property of these quantum systems, the so-called $R$-separation of
variables. 

In doing so, let us demonstrate that the non-stationary Schr\"odinger equation
\begin{equation}
  S[V_{1,2}]\psi_{1,2}(x,t)\equiv (i\partial_t +\partial^2_x -
  V_{1,2})\psi_{1,2}(x,t)=0 
\label{S12psi}
\end{equation}
with potentials as given in (\ref{V_1/2}), which is equivalent to the
intertwining (\ref{inS}), admits a separation of variables. In
fact, under the transformation 
\begin{equation}
\label{trafo}
  y=x/\rho(t) + \mu(t)\;,\qquad 
  \phi_{1,2}(y,t)=\sqrt{\rho(t)}\,e^{ig(x,t)}\psi_{1,2}(x,t)
\end{equation}
this Schr\"odinger equation becomes \cite{Bluman96} 
\begin{equation}
  i\rho^{2}(t)\partial_{t} \phi_{1,2}(y,t)=
  \left[-\partial_{y}^{2} + K'^{2}(y) \pm K''(y)\right]\phi_{1,2}(y,t)\;,
\label{SEiny}
\end{equation}
which is obviously separable in $y$ and $t$. Hence, the solutions of the
original problem (\ref{S12psi}) have the general form
$\psi(x,t)=\tilde{R}(y,t)Y(y)T(t)$ which 
is known as the $R$-separation of variables \cite{Miller77}. In other words,
for any pair of  
Schr\"odinger operators $S[V_{1,2}]$, which admits a first-order intertwining
relation (\ref{inS}) there exists a transformation (\ref{trafo}) to some new
coordinate in which the potentials become stationary and form the well-known
pair of stationary SUSY partner potentials \cite{Junker96}. As a consequence
nothing essentially new emerges from the first-order intertwining of the
non-stationary Schr\"odinger operator. This fact has also recently been
observed by Finkel et al.\ \cite{Finkel98}. 
Note that the transformation
associated with the special case $\rho(t)=1$ and $\mu(t)=vt$ with constant
velocity $v$ corresponds to the Galilei transformation. See, for example, the
textbook by Galindo and Pascual \cite{Galindo90}.

Here let us add that this $R$-separation of variables can
be related to the existence of symmetry operators for the two
Schr\"odinger operators in questions. First, we note that with the results
above one can directly verify the adjoint intertwining relation
\begin{equation}
  q^-S[V_1]=S[V_2]q^-
\label{inSadjoint}
\end{equation}
where
\begin{equation}
q^-\equiv(q^+)^\dagger=
  -\rho(t)\partial_x+K'\Bigl(x/\rho(t)+\mu(t)\Bigr)+
  \frac{i}{2}\left(\dot{\rho}(t)x-\rho^2(t)\dot{\mu}(t)\right).
\end{equation}
In other words, in analogy to the discussion of (\ref{inter+}) the
Schr\"odinger operator $S[V]$ and in particular its time-derivative
$i\partial_t$ can be viewed as self-adjoint operators on a suitable domain,
$(i\partial_t)^\dagger= i\partial_t$.
Then from (\ref{inS}) and (\ref{inSadjoint}) immediately follows
\begin{equation}
  \Bigl[S[V_1],q^+q^-\Bigr]=0\;,\qquad \Bigl[S[V_2],q^-q^+\Bigr]=0\;,
\end{equation}
which is a special case of (\ref{symmetry}). This discussion suggests that the
reverse conclusion may also be true. In fact we are able to prove the
following\\[2mm] 
{\it Theorem:} Any Schr\"odinger operator $S[V]$ (with real-valued potential
$V$) having a self-adjoint symmetry operator $R$, obeying $[S[V], R]=0$ and
being quadratic in $\partial_x$, admits the $R$-separation of variables for the
corresponding time-dependent Schr\"odinger equation and an intertwining
relation of first order with its superpartner.\\[2mm] 
For the proof of this theorem we start with the most general quadratic and
self-adjoint ansatz for the symmetry operator,
\begin{equation}
 R(x,t) = -\omega (t)\partial_{x}^{2} + i\{ \delta (x,t),\partial_{x} \} +
\zeta (x,t)\;, 
\label{R}
\end{equation}
obeying the symmetry condition
\begin{equation}
  [S[V], R]\equiv[ i\partial_{t} + \partial^{2}_{x} - V(x,t), R(x,t) ] = 0\;.
\label{SymmCond}
\end{equation}
Here, $\delta:{\mathbb R}\times{\mathbb R}\to{\mathbb R}$ and $\zeta:{\mathbb
R}\times{\mathbb R}\to{\mathbb R}$ are arbitrary real functions of position
and time, whereas $\omega:{\mathbb R}\to{\mathbb R}^+$ is, without loss of
generality, assumed to be an arbitrary strictly positive function of time
only. The coefficient functions of the symmetry operator (\ref{R}) are clearly
related with the potential $V$ due to the symmetry condition (\ref{SymmCond})
giving rise to
\begin{equation}
\delta (x,t) = \frac{1}{4}\,\dot\omega (t)x + \frac{1}{2}\,\nu (t)\;,\quad 
\zeta (x,t) = \Phi(z)+\frac{\dot{\omega}^2(t)}{16\,\omega(t)}\,x^2 
+\frac{\nu (t)\dot{\omega}(t)}{4\,\omega(t)}\,x\;,
\label{delta+zeta}
\end{equation}
where $\Phi$ is an arbitrary real function of the variable
\begin{equation}
 z=\frac{x}{\sqrt{\omega (t)}} - \int_0^{t}d\tau\,\frac{\nu (\tau)}
{\omega^{3/2}(\tau)}
\end{equation}
and $\nu$ is another arbitrary real function of time. In terms of these
functions the potential must take the form
\begin{equation}
  V(x,t) = -\frac{1}{8\omega (t)}\biggl(\ddot\omega (t) -
\frac{\dot\omega^{2}(t)}{2\omega (t)}\biggr) x^{2} -
\frac{1}{2\omega (t)}\biggl(\dot\nu (t) -
\frac{\nu (t)\dot\omega (t)}{2\omega (t)} \biggr) x +
\frac{\Phi (z)}{\omega (t)}\;.
\label{V}
\end{equation}
We note that the $R$-separation of variables for the Schr\"odinger equation
$S[V]\psi(x,t)=0$ becomes explicit by making the change of variables
\begin{equation}
  \phi(z,t)=\omega^{1/4}(t) \exp \left\{-i\left(\frac{\dot\omega (t)}
{8\,\omega (t)}\,x^{2}+\frac{\nu (t)}{2\,\omega (t)}\,x-
\frac{1}{4}\int_0^{t}d\tau\,\frac{\nu^{2}(\tau)}{\omega^{2}(\tau)}\right)
\right\}\psi(x,t) \;,
\end{equation}
which results in
\begin{equation}
\left(i\omega (t)\partial_{t} + \partial^{2}_{z} + \Phi (z) \right)
\phi (z,t) = 0\;.
\end{equation}
Let us remark that the symmetry operator $R$ of the form (\ref{R}) with
coefficient functions related by (\ref{delta+zeta}) is factorizable, i.e.\
$R=\tilde q^{+}\tilde q^{-}\geq 0$ with $\tilde q^{-}=(\tilde q^{+})^\dagger$. 
The supercharge $\tilde q^{+}$ has again the form 
(\ref{q+_2}) if the following substitutions are made:
\begin{equation}
 \rho (t) \rightarrow \omega^{1/2}(t)\;,\qquad
 \dot{\mu} (t) \rightarrow -\frac{\nu (t)}{\omega^{3/2} (t)}\;,\qquad
 y \rightarrow z\;,\qquad
 K'\,^2 +K'' \rightarrow \Phi\;. 
\end{equation}

Thus we conclude that if some non-stationary Schr\"odinger operator $S[V]$
has a symmetry characterized by a second-order operator of the form (\ref{R})
it admits the $R$-separation of variables and the superpartner potential of
$V$ can also be constructed. The non-stationary Schr\"odinger operator
associated with this superpartner has again a positive symmetry operator,
which in this case is given by $\tilde q^{-}\tilde q^{+}$. 

\subsection{The Fokker-Planck equation}
In this subsection we will now consider the most general first-order
intertwining relations allowing to construct from known solutions of one
Fokker-Planck equation solutions of its corresponding superpartner. In
order to keep as close as possible to the previous discussion we will first
transform the Fokker-Planck equation into a diffusion (or imaginary-time
Schr\"odinger) equation. Indeed, it is well-known \cite{Kampen92} that the
Fokker-Planck equation $F[U]P(x,t)=0$ transforms under the substitution
$P(x,t)=\exp\{-U(x,t)/2\}\psi(x,t)$ into the diffusion equation
\begin{equation}
  D[V]\psi(x,t)\equiv\left(-\partial_t+\partial_x^2-V(x,t)\right)\psi(x,t)=0\;,
\label{D}
\end{equation}
where the potential $V$ is given in terms of the drift potential $U$ in the
following way
\begin{equation}
  V(x,t)=\frac{1}{4}\,U'^{\,2}(x,t)-\frac{1}{2}\,U''(x,t)
           -\frac{1}{2}\,\dot{U}(x,t)\;.
\label{VvsU}
\end{equation}
Note that the diffusion operator $D[V]$ can be obtained from the corresponding
Schr\"odinger operator $S[V]$ via a Wick rotation
$i\partial_t\to-\partial_t$. There are, however, essential differences from
the physical point of view as the solution $P$ of the Fokker-Planck equation
should represent a probability distribution. As a consequence
$\psi(x,t)=\exp\{U(x,t)/2\}P(x,t)$ is also required to be real and positive.

We are now interested in finding the most general first-order intertwining
operator $q^+$ as given in (\ref{q+}) obeying the relation
\begin{equation}
  D[V_1]q^+=q^+D[V_2]\;\qquad \Longleftrightarrow \qquad 
  F[U_1]\,e^{U_1/2}q^+e^{-U_2/2}=e^{U_1/2}q^+e^{-U_2/2}\,F[U_2]\;.
\end{equation}
Here we closely follow the derivations of the previous subsection and the
result is identical to that given in (\ref{V1andV2:1}) after Wick
rotation. Again  
we may set $\alpha\equiv 0$ as it can be absorbed in $\chi$. Furthermore,
as noted above, the intertwining operator
$q^+=e^{i\beta}\rho(\partial_x+\chi')$ 
should map a positive solution of
$D[V_2]\psi_2=0$ into a positive solution of $D[V_1]\psi_1=0$,
i.e.\ $\psi_1=q^+\psi_2\geq 0$. This leads us
to the conclusion that $\beta$ must vanish identically and $\chi$ has to be
real-valued. Hence, we are left with 
\begin{equation}
  \begin{array}{l}
   V_1(x,t)=\chi'\,^2(x,t)+\chi''(x,t) + \dot{\chi}(x,t)
   -\dot{\rho}(t)/\rho(t)\;,\\[2mm]
   V_2(x,t)=\chi'\,^2(x,t)-\chi''(x,t) + \dot{\chi}(x,t)\;,
  \end{array}
\end{equation} 
giving rise, cf.\ (\ref{VvsU}), to the following form of the drift potentials
\begin{equation}
  \begin{array}{l}
  U_1(x,t)=2\,\ln \rho(t) - 2\,\chi(x,t)\;,\\[2mm]
  U_2(x,t)=2\,\chi(x,-t)\;.
  \end{array}
\label{U1andU2}
\end{equation}

In contrast to our result in the previous subsection we find here that $\chi$
has to be a real-valued function which, however, is a function of the two
independent variables $x$ and $t$. Therefore, the Fokker-Planck case can, in
contrast to the Schr\"odinger case, not be transformed into a stationary
problem. This seems to be related to the fact that the diffusion
operator is, in contrast to the Schr\"odinger operator, not
self-adjoint. Consequently, the adjoint intertwining relation (\ref{inter+})
and the resulting symmetries (\ref{symmetry}) do not exist. Physically, this
is due to the time-reversal symmetry, which exists in the case of the
Schr\"odinger equation \cite{Galindo90} but not for the Fokker-Planck and
diffusion equation. This is also explicated in the above result
(\ref{U1andU2}) which shows that the two SUSY-partner drift potentials are
essentially related via a time reversal. The additional difference in an
overall sign between them is already known from the stationary case
\cite{Junker96}.  

\section{Second-order intertwining for stationary potentials:
particular solutions and symmetry operators}
Recently higher-order generalizations of the SUSY-intertwining relations for
stationary Schr\"odinger operators have been investigated to some extend
\cite{ais,acdi}. 
In the 1-dimensional case it was shown \cite{acdi} that in general
the intertwining of a pair of stationary Schr\"odinger operators by a
second-order differential operator cannot be reduced
to two consecutive transformations with some intermediate self-adjoint
Hamiltonian. In the 2-dimensional case  
second-order intertwining operators also allow to intertwine pairs
of standard stationary Schr\"odinger Hamiltonians with scalar potential
\cite{ain1-3}. All of these self-adjoint Hamiltonians have a hidden symmetry,
cf.\ (\ref{symmetry}), characterized by a differential operator.

In the remaining part of this paper we will investigate the intertwining of a
pair of Schr\"odinger operators $S[V_1]$ and $S[V_2]$ by second-order
intertwining operators of the form: 
\begin{equation}
  q^{+}(x,t) = G(x,t)\partial_{x}^{2} - 2F(x,t)\partial_{x} + B(x,t)\;.
\label{old18}
\end{equation}
As in the first-order case it can be shown that the inclusion of an additional
term being of first order in $ \partial_{t} $ leads to the trivial situations
where the difference $V_1-V_2$ depends on the time $t$ only, cf.\
footnote \ref{footnote1}. Furthermore, from the intertwining relation
(\ref{inS}) with above $q^+$ one can conclude that the function $G$ may not
depend on $x$ and similar to the discussion in the previous section it is even
possible to exclude a phase. In other words, without loss of generality we
have $G(x,t)\equiv g(t)$, which should not be confused with $g$ used
previously, and consider from now on an intertwining operator of the
form 
\begin{equation}
  q^{+}(x,t) = g(t)\partial_{x}^{2} - 2F(x,t)\partial_{x} + B(x,t)\;.
\label{18}
\end{equation}

We are unable to find the general analytic solutions of the intertwining
relation (\ref{inS}) with $q^{+}$ as given above. Therefore, in this  
and the next section we will construct only some particular solutions of
interest. More precisely, in this section we shall look for the solutions
of the intertwining relation (\ref{inS}) for the case where both potentials
$ V_1$ and $V_2$ are stationary, i.e.\ do not depend on $t$.
It is evident that one class of solutions is already known from
\cite{acdi}. Assuming a supercharge $ q^{+} $ 
whose coefficient functions are real and do not depend on $t$, it follows that
the corresponding solutions of (\ref{inS}) will coincide with those of
the stationary intertwining relations $(-\partial^2_x + V_1(x))q^{+}(x) =
q^{+}(x)(-\partial^2_x + V_2(x))$ which can be found in \cite{acdi}.
Here we are interested in more general solutions of (\ref{inS}) with
operators $ q^{+}$ depending manifestly on $t$. That is, we will search for
particular solutions of the intertwining relation
\begin{equation}
  (i\partial_t -H_1)q^+(x,t)=q^+(x,t) (i\partial_t -H_2)\;
\label{inH}
\end{equation}
with standard stationary Hamiltonians $H_{1,2}=-\partial_x^2+V_{1,2}(x)$ but
an explicit time-dependent intertwining operator.

\subsection{Systems with symmetry operators of third order}
A first suitable ansatz with simple $ t$-dependence in (\ref{18}) is
\begin{equation}
  q^{+}(x,t) = M^{+}(x) + A(t)a^{+}(x)\;, 
  \label{20}
\end{equation}
where we have set
\begin{equation}
  M^{+}(x)\equiv \partial_{x}^{2} - 2f(x)\partial_{x} + b(x)\;,\qquad
  a^{+}(x)\equiv \partial_{x} + W(x)\;. \label{20a}
\end{equation}
Here all functions besides $A$ are considered to be real valued\footnote{We
  also assume $A\not\equiv 0$, as the case $A\equiv 0$ has already been
  studied in \cite{acdi}.}.  
With this ansatz the intertwining relation (\ref{inH}) results in the
condition
\begin{equation}
  i\dot Aa^{+} = H_1M^{+} - M^{+}H_2 + A \left(H_1a^{+} - a^{+}H_2 \right)\;.
\label{22}
\end{equation}
Equating the coefficient of identical powers in $\partial_{x}$ on both sides of
(\ref{22}) one obtains the following system of equations:
\begin{equation}
  \begin{array}{l}
i\dot A=2\,\tilde m + 2\,m\, A\;,\\[2mm]
H_1M^{+} - M^{+}H_2 = 2\,\tilde m\, a^{+}\;,\\[2mm]
H_1a^{+} - a^{+}H_2 = 2\, m\, a^{+}\;,
  \end{array}\label{25}
\end{equation}
with real constants $ \tilde m$ and $m$. For our further discussion we will
consider the two cases $m\neq 0$ and $m=0$ separately.\\

\noindent a) Solutions for $ m\neq 0 $:\\
In this case the first equation in (\ref{25}) immediately leads to
$A(t) = m_0e^{-2imt} - \tilde m/m$, $m_0\in{\mathbb R}$, and results in an
intertwining operator of the form
\begin{equation}
  q^{+}(x,t) = \partial_{x}^{2} - \biggl(2f(x) + \frac{\tilde m}{m}
\biggr)\partial_{x} + b(x) - \frac{\tilde m}{m}\,W(x) + m_0e^{-2imt}a^{+}(x)\;.
\end{equation}
It is obvious that without loss of generality we may set $\tilde m = 0$
as a non-vanishing $\tilde{m}$ may always be absorbed via a proper
redefinition of $f$ and $b$. As a consequence, the second relation in
(\ref{25}) leads to a second-order intertwining between $H_1$ and $H_2$.  
This has already been considered in \cite{acdi} and it was found that 
the potentials $V_1$, $V_2$ and the function $b$ can be expressed in terms of
$f$ and two arbitrary real constants $a$ and $d$:
\begin{equation}
  \begin{array}{l}
\displaystyle
V_{1,2}(x) = \mp 2 f'(x) + f^{2}(x) + \frac{f''(x)}{2f(x)} 
         - \frac{f'^{\, 2}(x)}{4f^{2}(x)}- \frac{d}{4f^{2}(x)} - a\;,\\[2mm]
\displaystyle
b(x) = -f'(x) + f^{2}(x) - \frac{f''(x)}{2f(x)} 
       + \frac{f'^{\, 2}(x)}{4f^{2}(x)} + \frac{d}{4f^{2}(x)}\; .
  \end{array}\label{27}
\end{equation}
Finally we note that the last equation in (\ref{25}) coincides with the usual
first-order intertwining relation of SUSY quantum mechanics between the
Hamiltonians $H_1$ and $ H_2 + 2m$. Therefore, the potentials can be expressed
in terms of the SUSY potential $W$ as follows \cite{Junker96}:
\begin{equation}
V_1(x) = W^{2}(x) + W'(x)\;, \qquad
V_2(x) = W^{2}(x) - W'(x) - 2m \;.
\label{28}
\end{equation}
Comparing this with (\ref{27}) we conclude that $W(x) = -2f(x) - mx$ and 
$f$ must satisfy the so-called Painleve IV equation:
\begin{equation}
f'' = \frac{f'^{\, 2}}{2f} + 6f^{3} + 8mxf^{2} + 2 (m^{2}x^{2}-m+a)f 
      + \frac{d}{2f}\;.
\label{29}  
\end{equation}

As an example we mention here that for $d=-a^2$ it can be shown
that any $f$ obeying the generalized Riccati equation $f'=-2f^2-2mxf-a$ is also
a solution of (\ref{29}). The general solution of this generalized Riccati
equation and the corresponding potentials (\ref{28}) are discussed in detail
in \cite{Junker98}, which also shows their connection to the most general SUSY
partners of the harmonic-oscillator potential. Note that cases with $d<0$
are reducible ones \cite{acdi}. 

We have already discussed in the previous section that the intertwining
relation (\ref{inH}) and its adjoint give rise to the symmetry operators
$q^{+}q^{-}$ and $q^{-} q^{+}$ for $(i\partial_{t} - H_{1})$ and
$(i\partial_{t} - H_{2})$, respectively. For the present case of stationary
potentials one can use results of \cite{acdi} to show that these symmetry
operators (defined up to a square polynomial in $H_{1,2}$) have the explicit
form 
\begin{equation}
  \begin{array}{l}
R_1(x,t) = e^{2imt}M^{+}(x)a^{-}(x) + e^{-2imt}a^{+}(x)M^{-}(x)\;,\\[2mm]
R_2(x,t) = e^{2imt}a^{-}(x)M^{+}(x) + e^{-2imt}M^{+}(x)a^{-}(x)\;,
  \end{array}\label{30}
\end{equation}
where $a^-=(a^+)^\dagger$ and $M^-=(M^+)^\dagger$. Let us note that these
results have also been found recently by Fushchych and Nikitin \cite{nikitin}
using a different approach.\\

\noindent b) Solutions for $m=0$:\\
In this case it follows from (\ref{25}) that $ A(t) = -2i\tilde mt$
where, without loss of generality, we have set the integration constant $A(0)$
to zero. The remaining two conditions of (\ref{25}) lead to 
\begin{equation}
  \begin{array}{l}
\displaystyle
  V_{1,2}(x) = W^{2}(x) \pm W'(x)\;,\qquad
  f(x) = n - \frac{1}{2}\,W(x)\;,\\[2mm]
\displaystyle
  b(x) = \frac{1}{2}\left(W'(x) - W^{2}(x)\right) -2nW(x) - \tilde{m} x\;,
  \end{array}\label{31}
\end{equation}
where $W$ is a solution of the Painleve II equation
\begin{equation}
  W'' = 2W^{3} + 4\tilde mxW + k
\label{32}
\end{equation}
and $ n, k $ are real constants. Here we note that an additional integration
constant appearing on the right-hand side of the last relation in (\ref{31})
has been set to zero as it can always be absorbed via a proper
redefinition of the independent variable $x$.

Again let us mention that for $k=-4\tilde{m}$ a particular solution of this
equation reads $W(x)=1/x$ giving rise to inverse-squared potentials
$V_{1,2}$. On the other hand, for $k=2\tilde{m}$ a solution of the ordinary
Riccati equation $W'=W^2+k\,x$ is also a solution of (\ref{32}). Note that
such solutions can be expressed in terms of Bessel functions \cite{Ince}.

As in the previous case the intertwining relation (\ref{inH}) leads to
symmetry operators, which can also be reduced to third-order differential
operators: 
\begin{equation}
  \begin{array}{l}
R_1 = M^{+}M^{-} - H_1^{2} + 2i\tilde m t ( M^{+}a^{-}-a^{+}M^{-}) 
      + 4\tilde m^{2}t^{2}H_1\;,\\[2mm]
R_2 = M^{-}M^{+} - H_2^{2} + 2i\tilde m t (a^{-} M^{+}-M^{-}a^{+}) 
      + 4\tilde m^{2}t^{2}H_2\;.
  \end{array}\label{33}
\end{equation}
However, in addition to that, because of $m=0$, we can even construct another
pair of third-order operators commuting with $ (i\partial_{t} -
H_{1,2})$. Indeed, the last two equations of (\ref{25}) can be rearranged 
as follows
\begin{equation}
 [H_1, M^{+}a^{-} ] = 2\tilde m H_1\;,\qquad
 [H_2, a^{-}M^{+} ] = 2\tilde m H_2
\label{34} 
\end{equation}
and directly lead to
\begin{equation}
[ i\partial_{t} - H_1, iM^{+}a^{-} + 2\tilde mtH_1 ] =
[ i\partial_{t} - H_2, ia^{-}M^{+} + 2\tilde mtH_2 ] = 0\;,
\end{equation}
Hence, besides those given in (\ref{33}), we have found an additional
hermitean pair of symmetry operators:
\begin{equation}
  \begin{array}{l}
\tilde R_1 = i\left(M^{+}a^{-} - a^{+}M^{-}\right) + 4\tilde m t H_1
           = \dot{R}_1/2\tilde{m}\;,\\[2mm]
\tilde R_2 = i\left(a^{-}M^{+} - M^{-}a^{+}\right) + 4\tilde m t H_2
           = \dot{R}_2/2\tilde{m}\; .
  \end{array}\label{35}
\end{equation}
Whereas these operators have already been given in \cite{nikitin} the previous
pair (\ref{33}) is a new, independent one, which does not commute with
(\ref{35}).   

\subsection{Systems with symmetry operators of fourth order}
As a second ansatz for the intertwining relation (\ref{inH}) we consider an
operator with two coefficient functions depending manifestly on time:
\begin{equation}
  q^{+}(x,t) = \theta (t) M^{+}(x) + i\lambda(t)\,x\,a^{+}(x) \;,
\label{36}
\end{equation}
where $\theta$ and $\lambda$ are assumed to be real-valued functions and the
operators $ M^{+}$ and $a^{+}$ are of the same form as in (\ref{20a}). For this
ansatz the intertwining relation (\ref{inH}) leads to
\begin{equation}
  \begin{array}{l}
\displaystyle
\dot{\theta} =-2\lambda\;,\qquad
\dot{\lambda} = \beta\theta\;,\\[3mm]
\displaystyle
2M^{+}   = xa^{+}H_2 - H_1 x a^{+}\;,\qquad
\beta\, x\, a^{+} = M^{+}H_2 - H_1M^{+}\; .
  \end{array}\label{38}
\end{equation}
The first two equations, containing the real constant $\beta$, can easily be
integrated and yield, for $\beta>0$, oscillatory solutions. 
Furthermore, with the above conditions the
potentials $ V_{1,2}$ and the functions $W$ and $b$ can exclusively be
expressed in terms of the function  $f$:
\begin{equation}
  \begin{array}{rl}
W(x) =&\displaystyle -2f(x) + \frac{c}{x}\;,\qquad 
b(x) = f'(x) + 2f^{2}(x) - V_2(x) +(\beta/4) x^{2} + a_{0}\;,\\[2mm]
V_{1,2}(x) =&\displaystyle \mp 2 f'(x) + 4 f^{2}(x) + \frac{2(1-2c)f(x)}{x} +
\frac{\beta}{8}\,x^{2}+\frac{4(c-1)f(x)}{x^2}\\
& \displaystyle - \frac{2}{x^{2}}
\int_{x_0}^{x}d z \left[f(z)-2zf^{2}(z)\right]\,+a_0\;,
  \end{array}\label{39}
\end{equation}
where $a_0$, $c$ and $x_0$ are some real constants and the function $f$ must
satisfy the non-linear third-order differential equation:
\begin{equation}
-2fV_2'+4bf'+f'''+4f'^2+4ff''+2\beta x f-\beta c+\beta/2=0\\
\label{40}
\end{equation}
This equation is certainly too complicate to solve for arbitrary values of the
constants $\beta$ and $c$. However, for $c=0$ and an arbitrary $d$ one can
show that all $f$'s satisfying the  Riccati equation
\begin{equation}
  f'(x) = 2f^{2}(x) - (\beta/4)x^{2} + d
\label{41}
\end{equation}
are also solutions of (\ref{40}). Such solutions correspond to
Hamiltonians obeying the intertwining relation $H_1a^{+} = a^{+}H_2$ in
addition to (\ref{38}).

It is straightforward to show that there are no other values\footnote{We
  exclude here the case $\beta =0$ which corresponds to the intertwining of
  $H_{1,2}$ by $ M^{+}$.} 
of the constant $c$ for which the Hamiltonians $ H_{1,2} $ are
intertwined by operators of first or second order. Since solutions of
eq.\ (\ref{41}) do not exhaust the solutions of (\ref{40}) we believe
that, at least in some cases, operators $ q^{+}$ of the form
(\ref{36}) intertwine the Schr\"odinger operators 
$S[V_{1,2}]$ but not the Hamiltonians $H_{1,2}$.

\subsection{Reducible and irreducible second order intertwining.}
The natural question we will now consider concerns the reducibility
of the second-order intertwining operator (\ref{18}) to a pair of consecutive
first-order operators $a_{1}^{+}$ and $a_{2}^{+}$. 
We again assume that the real potentials $V_{1,2}=V_{1,2}(x)$ are stationary
but the intermediate real potential $V$ may depend on time as well and to
which we may always add some arbitrary time-dependent function
$\Delta=\Delta(t)$. In other words, we 
consider the following chain of Darboux-type transformations between the
Schr\"odinger operators:
\begin{equation}
  S[V_1] \quad\stackrel{a^+_1}{\longrightarrow}\quad S[V]
  \quad\longrightarrow\quad  S[V+\Delta] 
  \quad\stackrel{a^+_2}{\longrightarrow}\quad S[V_2]\;.
\label{42}
\end{equation}
As we have shown in Section 2 a first-order intertwining for $S[V_1]$ and
$S[V_2]$ implies the existence of second-order symmetry operators, that is,
\begin{equation}
[S[V_1], a_{1}^{+}a_{1}^{-}] = [S[V_2], a_{2}^{-}a_{2}^{+}] = 0\;.
\end{equation}

Let us first consider that case where $a_{1}^{+}a_{1}^{-}$ is a trivial
symmetry operator, i.e.\ $a_{1}^{+}a_{1}^{-}=H_1+{\rm const}$. Then the
intertwining operators $a_{1}^{\pm}$ being of the form (\ref{q+})
with superpotential (\ref{chi}) can be simplified to  
\begin{equation}
 \rho (t)\equiv 1\;, \quad g(x,t)\equiv g(t)\;, \quad h(x,t)\equiv h(x)\;. 
\end{equation}
Thus the operators $a_{1}^{\pm}$ do not depend on time, 
$a_{1}^{\pm}=\pm\partial_{x}+h'(x)$. The same argumentation applies
to the operators $a_{2}^{\pm}$ and $H_2$. Evidently, in this case the
Hamiltonians $ H_{1,2}$ belong to the class considered in \cite{acdi}, where
it has been shown that reducible as well as irreducible second-order
intertwining operators exist.

In the case of a non-trivial symmetry operator $a_{1}^{+}a_{1}^{-}$ the
potential $V_1$ is of the form (see \cite{nikitin} and references therein)
\begin{equation}
V_1(x) = k_{0}+k_{1}x+k_{2}x^{2}+\frac{k_{3}}{(x+k_{4})^{2}}\;.
\end{equation}
It is evident that this class of potentials does not exhaust
all solutions of intertwining relations (\ref{inH}) with ans\"atze
(\ref{20}) or (\ref{36}) for $q^{\pm}$.

The conclusion drawn from this discussion is, that the intertwining relations
investigated in this section have both reducible and irreducible solutions of
second order in $\partial_x$.

\section{Second-order intertwining for non-stationary potentials}
In order to consider the intertwining of Schr\"odinger operators
with a non-stationary potential, it is useful to simplify the general form
(\ref{18}) of the intertwining operator $q^{+}$ a little further.
Indeed, by inspecting the general intertwining relation (\ref{inS}) with
$q^{+}$ as given in (\ref{18}) one finds the condition 
\begin{equation}
  \mbox{Im}\, F(x,t) = \frac{1}{4}\,\dot{g}(t)\,x + g_{1}(t)
\end{equation}
with an arbitrary real function $g_{1}$. After multiplying both
sides of (\ref{inS}) with
\begin{equation}
 g^{-1/4}(t)\exp\left[i\left(\frac{\dot{g}(t)x^{2}}{8g(t)} +
\frac{g_{1}(t)x}{g(t)} - \int_0^{t}dt'\,\frac{g_{1}^{2}(t')}{g^{2}(t')}
\right)\right]
\label{4.2} 
\end{equation}
and some algebra we obtain a new intertwining relation, 
\begin{equation}
\left( i\partial_{t}+2i\left(\frac{\dot gx}{4g(t)}+\frac{g_{1}}
{g}\right)\partial_{x}-\tilde H_1\right) \tilde q^{+} =
\tilde q^{+} \left( i\partial_{t}+2i\left(\frac{\dot{g}x}{4g}
+\frac{g_{1}}{g}\right)\partial_{x}-\tilde H_2 \right) \;.
\label{4.3}
\end{equation}
which, of course, is  equivalent to (\ref{inS}). Here the "new supercharge"
and "new Hamiltonians" are given by
\begin{equation}
  \begin{array}{ll}
\tilde q^{+}&\displaystyle
= g\partial_{x}^{2}-2(\mbox{Re}\, F)\partial_{x}+
B+\frac{i}{4}\,\dot g-\frac{1}{g}\left(\frac{\dot gx}{4}
+g_{1}\right)^{2}-\frac{2iF}{g}\left(\frac{\dot gx}{4}+g_{1}\right)\\[2mm]
&\displaystyle\equiv
g\partial_{x}^{2}-2(\mbox{Re}\, F)\partial_{x}+\tilde B 
  \end{array}
\label{4.4}
\end{equation}
and
\begin{equation}
\tilde H_{1,2}=H_{1,2}+\frac{1}{8}\left(\frac{\ddot g}{g}
-\frac{\dot g^{2}}{2g^{2}}\right)x^{2}
+\frac{1}{2}\left(\frac{g_{1}\dot g}{g^{2}}-\frac{2\dot g_{1}}{g}\right)x\;,
\label{4.5}
\end{equation}
respectively. 

Now, in a second step we make a non-local transformation \cite{Bluman96} of
the independent variables $(x,t)\mapsto(y,\tau)$ with
\begin{equation}
\tau=\int_0^{t}dt'\,\frac{1}{g(t')}\;,\qquad
y= g^{-1/2}(t)x-2\int_0^{t}dt'\,g_{1}(t')g^{-3/2}(t')\;,
\end{equation}
which brings the intertwining relation (\ref{4.3}) into the form
\begin{equation}
  \begin{array}{l}
\left( i\partial_{\tau}+\partial_{y}^{2}-g\tilde V_1\right) 
\left( \partial_{y}^{2}-2g^{-\frac{1}{2}}(\mbox{Re}\, F)\partial_{y}
  +\tilde B\right)=\\~~~~~~~~~~~~~~~~~~~~~~~~~~~~
\left( \partial_{y}^{2}-2g^{-\frac{1}{2}}(\mbox{Re}\, F)\partial_{y}
+\tilde B\right)
\left( i\partial_{\tau}+\partial_{y}^{2}-g\tilde V_2 \right)\;,
  \end{array}\label{4.6}
\end{equation}
where 
\begin{equation}
\tilde{V}_i(y,\tau)= V_i(x,t)+\frac{1}{8}\left(\frac{\ddot g}{g}
-\frac{\dot g^{2}}{2g^{2}}\right)x^{2}
+\frac{1}{2}\left(\frac{g_{1}\dot g}{g^{2}}-\frac{2\dot g_{1}}{g}\right)x\;.
\label{4.6a}
\end{equation}
This shows that, without loss of generality, we could have chosen from the
very beginning the operator $q^{+}$ to be of the form
\begin{equation}
  q^{+}(x,t)=\partial_{x}^{2}-2f(x,t)\partial_{x}+b(x,t)+ic(x,t)\;,
\label{4.1}
\end{equation}
with all coefficient functions being real-valued.

\subsection{Intertwining of a non-stationary Hamiltonian with a stationary one}
In Section 2 we have only briefly mentioned that a first-order intertwining
between Sch\"odinger operators for a non-stationary potential $V_1$ and a
stationary one $V_2$ is possible. Here, because of the absence of the
$R$-separation of variables,  we will reconsider this aspect for
the case of a second-order intertwining operator of the general form
(\ref{4.1}) in more details.

The intertwining relation (\ref{inS}) with ansatz (\ref{4.1}) leads to the
following conditions for the coefficient functions:
\begin{equation}
  \begin{array}{l}
\dot{f}=c'\;,\qquad \dot{b}+c''+4cf'=0\;,\qquad f''-b'-V_2'+4ff'=0\;,\\
\dot{c}+2fV_2'-b''-4bf'-V_2''=0\;,\qquad V_1=V_2-4f'\;.
  \end{array}\label{4.7}
\end{equation}
It may easily be verified that a particular set of solutions of (\ref{4.7})
can be expressed in terms of two arbitrary real-valued functions $f_{1}$ and
$f_0$, which depend only on $x$ and $t$, respectively: 
\begin{equation}
\begin{array}{l}
\displaystyle
f(x,t)=\frac{1}{2}\,\frac{f'_{1}(x)}{f_{1}(x)+f_{0}(t)}\;,\qquad
c(x,t)=\frac{1}{2}\,\frac{\dot f_{0}(t)}{f_{1}(x)+f_{0}(t)}\;,\\[4mm]
\displaystyle
V_2(x)=\frac{1}{f_{1}^{\prime 2}(x)}
 \left( \lambda_{0}^{2}\sigma\delta
 +\frac{1}{2}\left(f_{1}'(x)f_{1}'''(x)
 -\frac{1}{2}\,f_{1}''^{\, 2}(x)\right)
 -\frac{\lambda_{0}^{2}}{4}\,f_{1}^{2}(x)\right)\;,\\[4mm]
\displaystyle
V_1(x,t)=V_2(x)-4f'(x,t)\;,\qquad
b(x,t)=f'(x,t)+2f^{2}(x,t)-V_2(x)\;.
\end{array}\label{4.8}
\end{equation}
Recall that $b$, $c$, $f$ and $V_1$ are functions of $x$ and $t$ whereas $V_2$
is assumed to be stationary, i.e.\ independent of time $t$. This latter
assumption actually leads to the explicit form 
$f_{0}(t)=\sigma\exp{(\lambda_{0}t)}+\delta\exp{(-\lambda_{0}t)}$ 
with arbitrary constants $\sigma$, $\delta$ and $\lambda_{0}$. For this 
case, the symmetry operators $ R_i$ associated with the Schr\"odinger
operators $S[V_i]$ read
\begin{equation}
  \begin{array}{l}
R_1=q^{+}q^{-} =H_1^{2} + \frac{1}{4}\,\lambda_{0}^{2} - 2\dot{c}
       -4ic'\partial_{x} - 2ic''\;,\\[2mm]
R_2=q^{-}q^{+} = H_2^{2} + \frac{1}{4}\,\lambda_{0}^{2}\;,
  \end{array}\label{4.9}
\end{equation}
with $H_i=-\partial_x^2+V_i$. Note that $R_1$ explicitly depends on time,
whereas $R_2$ is stationary, a fact, which helps to show that the transformed
function 
$\psi_1=q^+\psi_2$ of a normalized solution of $S[V_2]\psi_2(x,t)=0$ is a
normalizable solution of $S[V_1]\psi_1(x,t)=0$. Indeed, because of the
linearity of the Schr\"odinger operator and the stationarity of $V_2$, we may
set 
$\psi_2(x,t)\equiv\psi_{2}(x,t;E) = e^{-iEt}\varphi_{E}(x)$ with
$\varphi_{E}$ being an eigenstate of $H_2$ associated with eigenvalue $E$.
Therefore, we have
\begin{equation}
\left( \psi_{1}(x,t;\tilde{E}), \psi_{1}(x,t;E) \right)\equiv
\left( q^+\psi_{2}(x,t;\tilde{E}), q^+\psi_{2}(x,t;E) \right)=
\left(E^{2}+\lambda_{0}^{2}/4\right)\delta_{\tilde E\, E}\;.
\label{norm}
\end{equation}

Finally, we mention that in the particular case of a constant potential 
$V_2$ (without loss of generality we may set $V_2\equiv 0$) the function
$f_{1}$ has to be a solution of the linear fourth-order differential equation
with constant coefficients (see eq.\ (\ref{4.8}))
\begin{equation}
f_{1}'''' - \lambda_{0}^{2}f_{1} = 0\;,
\label{4.11}
\end{equation}
whose general solution may easily be constructed by standard methods and will
be an arbitrary linear combination of the fundamental solutions
$\sin(\sqrt{\lambda_0}\,x)$, $\cos(\sqrt{\lambda_0}\,x)$,
$\sinh(\sqrt{\lambda_0}\,x)$ and $\cosh(\sqrt{\lambda_0}\,x)$.
Note that the partner potential 
$V_1(x,t)=-2\left[\partial^2_x\log(f_1(x)+f_0(t))\right]$
has the same scattering properties as $V_2\equiv 0$. That is, it is a
reflectionless potential. As particular example we may choose
$f_1(x)=\cosh(x)$ and $f_0(t)=\cosh(t)$ which leads to 
$V_1(x,t)= -2\cosh(x)\sinh(t)/(\cosh(x)+\cosh(t))^2$. 

\subsection{Intertwining of a non-stationary Hamiltonian with a time-dependent
  harmonic-oscillator Hamiltonian} 
In the remaining part of this section we will now consider the second-order
intertwining of Schr\"odinger operators associated with two non-stationary
Hamiltonians. In doing so we will limit ourselves to one of the simplest
non-stationary Hamiltonians, namely, that of a harmonic oscillator with a
time-dependent frequency and search for the class of non-stationary
potentials $V_1$ which are intertwined with this one.

As a starting point we choose the potential to be of the form
\begin{equation}
V_2(x,t) = -\frac{1}{8}\left( \frac{\ddot g(t)}{g(t)} -
\frac{\dot g^{2}(t)}{2g^{2}(t)} \right) x^{2}\;,
\end{equation}
which corresponds to the trivial potential $\tilde V_2\equiv 0$ in
(\ref{4.6a}) as we also set $ g_{1}\equiv 0 $ in the following discussion.
As a consequence we obtain from (\ref{4.6}) the intertwining relation:
\begin{equation}
\left( i\partial_{\tau}+\partial_{y}^{2}-\tilde U_1 \right)
\left( \partial_{y}^{2}-2\tilde f(y,\tau)\partial_{y}+\tilde B(y,\tau)
\right)=
\left( \partial_{y}^{2}-2\tilde f(y,\tau)\partial_{y}+\tilde B(y,\tau) \right)
\left( i\partial_{\tau}+\partial_{y}^{2} \right)\;,
\label{4.12}
\end{equation}
where 
\begin{equation}
\tilde U_1(y,\tau)= g(t)\tilde V_1(y,\tau)\;,\quad
\tilde f(y,\tau)= g^{-1/2}(t)\mbox{Re}\, F(x,t)
\end{equation}
are given in terms of the new variables $y=x/\sqrt{g}$ and 
$\tau=\int_0^{t}dt'\,g^{-1}(t')$. Taking into account the relations (\ref{4.8})
we have
\begin{equation}
  \begin{array}{l}
\tilde U_1(y,\tau)=-4(\partial_y\tilde f)(y,\tau)\;,\qquad
\tilde B(y,\tau)=(\partial_y\tilde f)(y,\tau) + 2\tilde f^{2}(y,\tau)+ 
  i\tilde c(y,\tau)\\[2mm]
\displaystyle
\tilde f(y,\tau)=\frac{1}{2}\,
\frac{(\partial_y\tilde f_{1})(y)}{\tilde f_{1}(y)+\tilde f_{0}(\tau)}\;,\qquad
\tilde c(y,\tau)=\frac{1}{2}\,
\frac{(\partial_\tau\tilde f_{0})(\tau)}{\tilde
  f_{1}(y)+\tilde f_{0}(\tau)}, 
  \end{array} \label{4.14}
\end{equation}
where $\tilde f_{0}(\tau)=\tilde\sigma e^{\tilde\lambda_{0}\tau}
+\tilde\delta e^{-\tilde\lambda_{0}\tau}$
and $ \tilde f_{1} $ is an arbitrary solution of (\ref{4.11}) with $\lambda_0$
and $x$ replaced by $\tilde\lambda_{0}$ and $y$, respectively. 
Returning back to original variables $ (x,t)$ and setting
$g(t)\equiv\rho^{2}(t),$ we find that the original potentials read
\begin{equation}
  \begin{array}{l}
\displaystyle
V_1(x,t) = -\frac{\ddot\rho (t)}{4\rho (t)}\, x^{2} -
2\left[\partial_x^2\log\left(
\tilde f_{1}\left(\rho^{-1}(t)x\right) \textstyle
+ \tilde f_{0}\left(\int_0^tdt'\,\rho^{-2}(t')\right)\right)\right]
\;,\\
\displaystyle
V_2(x,t) = -\frac{\ddot\rho (t)}{4\rho (t)}\, x^{2}\;,
  \end{array}\label{4.15}
\end{equation}
which may be compared with the result of Bluman and Shtelen \cite{Bluman96}
who derived new potentials being related to the free quantum system (in
the $y$-$\tau$ coordinates) via non-local transformations. Here we note that
the special case $\rho(t)=\cos (\omega t)$ leads to the ordinary (i.e.\
time-independent) harmonic oscillator case and the corresponding
transformation has been called Jackiw transformation.\footnote{See, for
  example, \cite{Jackiw80,Cai82} and \cite{Junker85} where more general
  transformations are given.} 

The present result is also closely related to the special case
$K'^2-K''=\mbox{const.}$ discussed in Section 2.1 where we have shown its
close connection with the R-separation of variables. Thus solutions for $V_1$
in (\ref{4.15}) can be related to a free quantum system characterized by
(\ref{SEiny}). 

\section{Concluding remarks}
In this paper we have considered the most general time-dependent first- and
second-order intertwinings of non-stationary Schr\"odinger operators. Whereas,
in the first-order case a complete discussion was possible, in the second-order
case only particular examples have been discussed. These, in turn, are not as
simple as those of the first-order case. 

In the main text we have concentrated on the implications of the intertwining
relations and did not take into account secondary aspects such as zero modes
and domain questions, which we are going to discuss now. As already mentioned
in the introduction, the intertwining may provide a new non-trivial solution
$\psi_1=q^+\psi_2$ of a given problem only in the case when $\psi_2$ does not
belong to the kernel of the intertwining operator $q^+$. Elements of the
kernel of $q^+$ lead to trivial solutions. On the 
other hand, $S[V_1]\psi_1=0$ may have non-trivial solutions which cannot be
obtained from solutions of $S[V_2]\psi_2=0$. In this case, $\psi_1$
necessarily belongs to the kernel of the adjoint intertwining operator $q^-$,
$q^-\psi_1=0$. Therefore, in order to find a complete set of solutions one has
to consider also the kernel of the intertwining operator and that of the
adjoint one. Again the first-order case allows a general discussion. The
result of our discussion in that case (Section 2) has led us to the general
form $q^+=\rho(\partial_x+h'+ig')$, where $h$ and $g$ are given in
(\ref{g_and_h}). As $q^+$ is a first-order differential equation the dimension
of its kernel is at most one-dimensional:
\begin{equation}
  \psi_2(x,t)\propto\exp\{-h(x,t)-ig(x,t)\}\;.
\end{equation}
The restriction of solutions to the linear
space of square-integrable functions then leads to the normalization condition
\begin{equation}
  1=\int dx |\psi_2(x,t)|^2=\int dy \exp\{-2 K(y)\}\;.
\end{equation}
Hence, the question of zero modes of $q^+$ or its adjoint is equivalent to a
discussion of broken versus unbroken SUSY of the corresponding
stationary problem in the $y$-coordinate \cite{Junker96}. The situation in the
case of 
second-order intertwining is a bit more complicated. However, it is clear that
the corresponding kernels are at most 
two-dimensional. In the particular case studied in Section 4.1 the kernel of
$q^+$ is obviously empty for a real non-vanishing parameter $\lambda_0$ as in
this case the symmetry operator $R_2=(q^+)^\dagger q^+$, cf.\ eq.\ (\ref{4.9}),
is strictly positive. 

Another aspect which we did not discuss in the main text is concerned with
domain questions. Indeed, one usually is interested only in square-integrable
solutions of the Schr\"odinger equation. There is no guarantee that the
intertwining operator maps a square-integrable solution into another
square-integrable one. This typically happens when the superpotential or, more
generally, the coefficient functions of the intertwining operator become
singular \cite{Junker96}. Hence, the rather general and abstract results
presented in the main text should be understood in the sense that whenever
there appear singular coefficient functions a careful analysis of such domain
questions is mandatory. For example, the case of inverse-square potentials
briefly mentioned in Section 3.1 b) is only well-defined if the domain
(Hilbert space) is defined by the space of square-integrable functions on the
positive half-line with proper (e.g.\ Dirichlet) boundary conditions at
$x=0$.

Finally, let us mention that it is also possible to extend our approach to
complex-valued potentials $V_{1,2}$, which recently have
attracted some attention \cite{Bender98}. The case of stationary complex
potentials with first-order intertwining has already been studied in
\cite{Cannata98,acdi98}. However, because of the absence of the reality
conditions, cf.\ eq.\ (\ref{h_and_g}), the first-order intertwining of such
potentials will not lead to the $R$-separation of
variables. Furthermore, the corresponding Schr\"odinger operators are no
longer self-adjoint and, therefore, no symmetry operators of the form $q^+q^-$
and $q^-q^+$ exist. This situation is similar to that of the Fokker-Planck and
diffusion equation discussed in Section 2.2. For the higher-order intertwining
the notion of irreducible transformations is lost in the case of complex
potentials. 

\section*{Acknowledgements}
This work was partially (M.I.\ and D.N.) supported by RFBR. One of us
(M.I.) is grateful to the DAAD for supporting a research visit to the 
University of Erlangen-N\"urnberg and also likes to thank the INFN for
supporting a subsequent visit to the University of Bologna.  Kind hospitality
of both institutions is also gratefully acknowledged.


\end{document}